\documentclass[11pt]{article}
\usepackage{amstext,amssymb}
\usepackage{latexsym,epsfig}

\newcommand{\ie}{{\em i.e., }}
\newcommand{\eg}{{\em e.g., }}

\newcommand{\remove}[1]{}

\newcommand{\openbox}{\leavevmode
  \hbox to.77778em{%
  \hfil\vrule
  \vbox to.675em{\hrule width.6em\vfil\hrule}%
  \vrule\hfil}}
\newcommand{\proofname}{Proof}

\begin{document}
\renewcommand{\textfraction}{0}

\title{The Viterbi Algorithm:  A Personal History}
 \author{\normalsize
G. David Forney, Jr.
 \\[-5pt]
\small MIT \\[-5pt]Ê
\small Cambridge, MA 02139 USA \\[-5pt] \small
\texttt{forneyd@comcast.net} }
\date{}
\maketitle
\thispagestyle{empty}
\begin{abstract}
The story of the Viterbi algorithm (VA) is told from a personal
perspective. Applications both within and beyond communications are
discussed. 
In brief summary, the VA has proved to be an extremely important algorithm
in a surprising variety of fields.
\end{abstract}Ê
\normalsize


\section{Introduction}

Andrew J. Viterbi is rightly celebrated as one of the leading
communications engineers and theorists of the twentieth century.  He has
received almost every professional award possible, including election not
only to the National Academy of Engineering (USA) but also to the National
Academy of Sciences (USA), where he chairs the Computer
Science section.  His award citations usually cite ``invention of the
Viterbi algorithm" as his most notable accomplishment.

On the other hand, Andy would be the first to tell you that other people
deserve much of the credit for recognizing its theoretical properties and
its practical attractiveness, and for extending its domain of
application.  He has often told this story himself (see, \eg  \cite{V90}).

Nevertheless, no one doubts that Andy's awards are entirely deserved, and
that their focus on the Viterbi algorithm (VA) is 
appropriate.  This article will attempt to explain why, by briefly
recounting the history of the VA.  It is a ``personal history," because
the story of the VA is so intertwined with my own history that I can
recount much of it from a personal perspective.

\section{Invention of the Viterbi algorithm}

The Viterbi algorithm was first presented in Andy's famous 1967 paper
\cite{V67a} to help prove an asymptotically optimum upper bound
on the error probability of convolutional codes, which had previously been
derived by Yudkin in the context of sequential decoding \cite{Y64}.  In
this paper, the VA is presented just as we understand it today.  This paper
introduces the important concept of \emph{survivors} (a term possibly
borrowed from tennis elimination tournaments), and shows that
only $q^K$ survivors need be retained to decode a convolutional code with
constraint length $K$ over the
$q$-ary field $GF(q)$.  Compared to a block
code with $q^K$ codewords, such a convolutional code is shown to have a
much better error exponent, particularly near
capacity.

Andy recalls in a 1999 interview \cite{M99} that

\begin{quote}
 ``the Viterbi algorithm
for convolutional codes \ldots came out of my teaching \ldots.  I found
information theory difficult to teach, so I started developing some tools.
\ldots  I wrote the first paper in March `66, but it wasn't published
until April `67. \ldots  At one point I was actually discouraged from
publishing the algorithm details.  Fortunately, one of the reviewers, Jim
Massey, encouraged me to include the algorithm. \ldots  Nobody thought
that it had any potential for practical value \ldots"
\end{quote}

It is clear from the paper that at this point Andy had no idea that the VA
was actually an optimum (maximum likelihood) decoder, nor that it was
potentially practical.  Indeed, the paper states that ``this decoding
algorithm is clearly suboptimal," and concludes: ``Although this algorithm
is rendered impractical by the excessive storage requirements, it
contributes to a general understanding of convolutional codes and
sequential decoding through its simplicity of mechanization and analysis"
\cite{V67a}.


\section{Discovery that the VA is optimum}

I believe that I received a copy of Andy's paper prior to publication,
probably via Jim Massey.  At that time I was working at Codex Corp., a
small start-up company aiming at practical applications of convolutional
codes.  Our primary focus was initially on threshold decoding, which was
the subject of Jim's doctoral thesis \cite{M63};  Jim was a
consultant.  Subsequently, we developed a sequential decoding system
\cite{WR61} for the Pioneer deep-space satellite program, which became the
first code in space \cite{CHIW98}.

I had been trying to understand why in practice convolutional codes were
generally superior to block codes, so I studied Andy's paper
with great interest.  I realized that the path-merging property of
convolutional codes could be depicted in what I called a \emph{trellis
diagram}, to contrast with the then-conventional tree diagram used in the
analysis of sequential decoding.  It was then only a small step to see that
the Viterbi algorithm was an exact recursive algorithm for finding the
shortest path through a trellis, and thus was actually an optimum trellis
decoder.  I believe that at that point I called Andy, and told him that he
had been too modest when he asserted that the VA was ``asymptotically
optimum."

These results were written up in a 1967 technical report \cite{F67}
for NASA Ames Research Center.  They were not published in journal form
until many years later, in \cite{F73} and \cite{F74}.

Shortly afterward, in a paper submitted in May 1968  \cite{O69},
Jim Omura observed that the VA was simply the standard
forward dynamic programming solution to maximum-likelihood decoding of a
discrete-time, finite-state dynamical system observed in memoryless
noise.  Beyond proving optimality in a different way, he thus made the
first connection between the VA and system and control theory.  It is
interesting to speculate whether the history of the VA would have been
different if it had simply been called ``dynamic programming" from the
beginning. 

At this point, none of us had recognized that the VA might be practical. 
Jim's paper concludes:  ``\ldots the decoding algorithm discussed here
grows exponentially in complexity with constraint length $\nu$ and is
therefore impractical for large $\nu$ \ldots."  More embarrassingly, in a
1970
\textsc{IEEE Spectrum} paper \cite{F70} describing practical coding
schemes for the space channel, I wrote:

\begin{quote}
Sequential decoding [is] the best-performing practical technique known for
memoryless channels like the space channel, and will probably be the
general-purpose workhorse for these channels in the future \ldots.

[The Viterbi algorithm] is competitive in performance with sequential
decoding for moderate error rates, but cannot achieve very low error rates
efficiently.  On the other hand, it [is] capable of extremely high speeds
(tens of megabits), where sequential decoders become uneconomic.  It
therefore may find application in high-data-rate systems with modest error
requirements, such as digitized television.
\end{quote}

\section{Recognition that the VA is practical}

Andy has always said that Jerry Heller was the first person to realize
that the VA might be practical.  Jerry simulated the performance of
short-constraint-length codes at the Jet Propulsion Laboratory (JPL) in
1968-69
\cite{H68, H69}, and found that with only a 64-state code he could
obtain a sizable coding gain, of the order of 6 dB.

In 1968, Andy, Irwin Jacobs, and Len Kleinrock incorporated Linkabit
Corp.\ in San Diego as a vehicle to pool their consulting efforts and to
obtain small government study contracts.  All kept their jobs as
professors.  In 1969, Jerry Heller was hired as Linkabit's first full-time
employee.  Linkabit obtained some small Navy and NASA contracts,
which enabled the construction of a VA prototype in 1969-70.  ``It was a
big monster filling a rack"
\cite{M99}.

The first IEEE Communication Theory Workshop in 1970 in St.\ Petersburg
became famous as the ``coding is dead" workshop, after Ned Weldon and other
speakers worried publicly that coding theory had come to a dead end.  But
what I remember best from that session is Irwin Jacobs standing up in the
back row, flourishing an integrated circuit (a 4-bit shift register, I
believe), and asserting that this represented the future of coding.  He
was quite right. (Unfortunately, by this time Codex had made a business
decision to get out of coding.)

By 1971, Linkabit had implemented a 2 Mb/s, 64-state Viterbi decoder.  In
a special issue on coding of the \textsc{IEEE Transactions on Communication
Technology} in October 1971, Heller and Jacobs
\cite{HJ71} discuss this decoder and many practical issues in careful
detail.  They compare the VA with sequential decoding, and conclude that
the VA will often be preferable because it can use quantized soft decisions
easily, and is less sensitive to channel and equipment variations.  In the
same issue, Cohen, Heller and Viterbi \cite{CHV71} describe a system using
orthogonal convolutional codes and the VA for asynchronous multiple-access
communications, and Viterbi \cite{V71} introduces generating-function
analysis techniques for the VA.


During the 1970s, through the leadership of Linkabit and JPL, the VA
became part of the coding standard for deep-space communication,
ultimately in a concatenated coding system with a Reed-Solomon (RS) outer
code.  Linkabit developed a relatively inexpensive and flexible VA chip,
and the VA became a nice little business for Linkabit.  It didn't hurt
that the inventor of the Viterbi algorithm was a Linkabit founder.  The VA
also began to be incorporated in many other communications applications.

In the early 1990s, JPL built a $2^{14}$-state ``Big Viterbi Decoder"
(BVD) with 8192 parallel add-compare-select (ACS) units, which operated at
a rate of the order of 1 Mb/s \cite{C92}.  As far as I know, the BVD
remains the biggest Viterbi decoder ever built.  

When the primary antenna
failed to deploy during the Galileo mission in 1992, JPL devised an
elaborate concatenated coding scheme involving a $2^{14}$-state rate-1/4
inner convolutional code and a set of variable-strength RS outer codes,
and reprogrammed it into the spacecraft computers.  This scheme was
able to operate within about 2 dB of the Shannon limit at a bit error
probability of less than $10^{-6}$, which was the world record prior to
the advent of turbo codes \cite{CHIW98}.

\section{The VA and intersymbol interference channels}

In the late 1960s, Codex turned its attention to the voiceband
modem business.  Our first-generation product
was a single-sideband (SSB) 9600 b/s modem with a so-called Class IV or $1-
D^2$ ``partial response."  About 1969, I recognized that the symbol
correlation that was thus introduced could be exploited by an \emph{ad
hoc} error correction algorithm, which was able to improve the noise
margin by about 2--3 dB.  This little decoder extended the commercial life
of this marginal-performance modem by perhaps a year or two.

It took me a while to understand that I had in fact invented a
maximum-likelihood sequence detector for this modem.  Over time, I
realized that this was nothing more than the Viterbi algorithm again,
streamlined for the $1 - D^2$ response.  This led to a 1972 paper
\cite{F72} that showed that the VA could be used as a maximum-likelihood
sequence detector for digital sequences in the presence of intersymbol
interference (ISI) and AWGN noise.  

Meanwhile, Jim Omura had recognized independently at UCLA that the VA
could be used on intersymbol interference channels, because of their
convolutional character \cite{O71}.  Indeed, a tantalizing hint in this
direction appears in a book review by Andy Viterbi in 1970 \cite{V70}. 
After visiting UCLA, Hisashi Kobayashi further developed this idea,
particularly for practical applications in partial response modems  and
magnetic recording \cite{K71a, K71b}.

The VA proved to be too complicated for general use as an equalizer on ISI
channels.  However, it stimulated many suboptimal approximations, and
analysis of its performance gave bounds on the best possible
performance of any sequence detector.

However, the VA did become standard in the related application of
high-density magnetic recording.  In so-called PRML systems
(``partial-response equalization with maximum-likelihood sequence
detection") \cite{ISW98}, the magnetic recording channel is first equalized
to a simple ``partial response" such as $1 - D^2$,
and the resulting sequence is then detected by the VA, or by a simplified
version thereof, as Kobayashi had envisioned \cite{K71a}. 
In retrospect, it seems possible that my little SSB modem
decoder was the first implementation of such a PRML scheme.

\section{Trellis-coded modulation}

After Gottfried Ungerboeck published his invention of trellis-coded
modulation in 1982 \cite{U82}, the VA became the workhorse
decoder for the next several generations of voiceband modems.  
Ungerboeck extended trellis coding to multilevel constellations by
constructing trellis codes in which each branch of the trellis represents
a subset of constellation symbols, rather than a single symbol.  By clever
constellation partitioning and attention to distances between subsets, he
was able to obtain coding gains in the bandwidth-limited regime
comparable to those that can be obtained in the power-limited
regime.  

For example, the V.32 modem (1986) used an 8-state trellis code to
obtain a coding gain of about 3.5 dB, while the later V.34 modem (1994)
 used 16 to 64-state trellis codes to obtain coding gains of
4.0 to 4.5 dB \cite{FBEM96}.

\section{Applications in mobile and broadcast communications}

The mobile communications channel is subject to fading, bursts, and
multiuser interference, and is a much more difficult medium than the
AWGN and linear Gaussian channels discussed above.  
The designers of second-generation (2G) cellular systems used every tool
available at the time (early 1990s) to provide reliable communication on
this difficult channel.

The CDMA system developed by Qualcomm uses a $2^8$-state, rate-1/3
convolutional code with interleaved 64-orthogonal modulation, and of course
a Viterbi decoder.  The TDMA system developed for GSM uses the VA
not only to decode a 16-state, rate-1/2 convolutional code, but
also for equalization.  A soft-output Viterbi algorithm (SOVA) is often
used in the latter application \cite{CHIW98}.

VA decoders are currently used in about one billion cellphones, which is
probably the largest number in any application.  However, the largest
current consumer of VA processor cycles is probably digital video
broadcasting.  A recent estimate at Qualcomm is that approximately
$10^{15}$ bits per second are now being decoded by the VA in digital TV
sets around the world, every second of every day \cite{P05}.

\section{General application to hidden Markov models}

In 1973, I wrote a tutorial paper on the Viterbi algorithm for the
\textsc{Proceedings of the IEEE} \cite{F73} that has turned out to be my 
most cited paper by far.  A recent search using Google Scholar shows 734
citations, far more than the 181 for my next-most-cited reference.

One of the main points of that paper was that the VA can be applied to any
problem that involves detecting the output sequence of a discrete-time,
finite-state machine in memoryless noise--- \ie to detection and pattern
recognition problems involving hidden Markov models (HMMs).  Of course,
decoding of convolutional codes and sequence detection on ISI channels
were the main applications discussed in that paper.

During the 70s and 80s, the VA became widely used
in a variety of pattern recognition problems that could be described by
HMMs, particularly for speech recognition;  see
\cite{R89}.  Here the VA is often used as the M-step of an EM algorithm,
which also adjusts HMM parameters.

Indeed, a recent search of IEEE Xplore shows that most
current IEEE references to the VA occur in such Transactions as
\textsc{Pattern Analysis and Machine Intelligence} or \textsc{Systems, Man
and Cybernetics},
rather than in \textsc{Communications} or \textsc{Information Theory}. 
It seems that everyone in these fields knows how to ``Viterbi the data."

Finally, in the past decade, the VA has become widely used in much more
distant fields such as computational biology, \eg to locate genes in DNA
sequences.  See for example \cite{HSF97}, with its ``Viterbi Exon-Intron
Locator" (VEIL).

\section{Related algorithms}

In the past decade, the development of the field of ``codes on graphs" and
their related decoding algorithms has led to a remarkable conceptual
unification of a variety of detection and estimation algorithms which have
been introduced under various names for various applications.

In his 1996 dissertation, generalizing the earlier work of Gallager
\cite{G63} and Tanner \cite{T81}, Niclas Wiberg
\cite{W96, WLK95} developed the generic ``sum-product" and ``min-sum"
decoding algorithms for cycle-free graphs which may include
both symbol (observable) and state (hidden) variables.  For trellis
graphs, he showed that these reduce to the BCJR algorithm
\cite{BCJR74} and an algorithm equivalent to the Viterbi algorithm,
respectively.  For capacity-approaching codes such as turbo codes and
low-density parity-check (LDPC) codes, the sum-product algorithm with an
appropriate schedule becomes the standard iterative decoding algorithm
that is normally used with such  codes.

Later authors (\eg \cite{AM00, KFL01}) have shown that the sum-product
algorithm is equivalent to Pearl's ``belief propagation" algorithm for
statistical inference on Bayesian networks;  the Baum-Welch or
``forward-backward" algorithm for inference with hidden Markov models; and
the Kalman smoother for linear Gaussian state-space models.

However, it is important to note that the min-sum algorithm
is a two-way ``backward-forward" algorithm.  The VA obtains the same
result with a ``forward-only" algorithm by storing a path history with
each survivor.  Of course, ``forward-only" is a key simplification,
particularly for real-time communications;  the min-sum algorithm would
never have been adopted in practice as widely as the VA has
been.\footnote{Interestingly, Ungerboeck discovered both the sum-product
and the min-sum algorithms for equalization applications in his thesis
\cite{U71};  however, he missed the forward-only version.}

\section{Conclusion}

The Viterbi algorithm has been tremendously important in communications. 
For moderately complex (not capacity-approaching) codes, it has proved to
yield the best tradeoff between performance and complexity both on
power-limited channels, such as space channels, and on bandwidth-limited
channels, such as voiceband telephone lines.  In practice, in these regimes
it has clearly outstripped its earlier rivals, such as sequential decoding
and algebraic decoding.  (However, it seems likely that it will be
superseded in many of its principal communications applications by
capacity-approaching codes with iterative decoding.)

Moreover, the VA has become a general-purpose algorithm for
decoding hidden Markov models in a huge variety of applications, from
speech recognition to computational biology. 

Andy Viterbi clearly did not envision the full import of the VA when
he first introduced it.  However, he and his colleagues at Linkabit and
Qualcomm were largely responsible for making it practical, and for driving
its widespread adoption in communications.  The history might have been
otherwise, but it wasn't.  In actual fact, no one deserves more credit for
this tremendously important invention than its actual inventor.

\section*{Acknowledgments}
I am very grateful for comments on drafts of this paper by Keith Chugg,
Dan Costello, Bob Gallager, Jim Massey, Jim Omura, Sergio Verd\'{u} and
Andy Viterbi.


{\small
\begin{thebibliography}{10}

\bibitem{AM00}
S. M. Aji and R. J. McEliece, ``The generalized distributive law,"
\emph{IEEE Trans.\ Inform.\ Theory}, vol.\ 46, pp.\ 325--343, Mar.\
2000.

\bibitem{BCJR74}
L. R. Bahl, J. Cocke, F. Jelinek and J. Raviv,
``Optimal decoding of linear codes for minimizing symbol error rate,"  
\emph{IEEE Trans.\ Inform.\ Theory}, vol.\ IT--20, pp.\ 284--287, Mar.\
1974.

\bibitem{CHV71}
A. R. Cohen, J. A. Heller and A. J. Viterbi, ``A new coding technique
for asynchronous multiple access communication," 
\emph{IEEE Trans.\ Commun.\ Tech.}, vol.\  COM--19, pp.\ 849--855, Oct.\
1971.

\bibitem{C92}
O. M. Collins, ``The subtleties and intricacies of building a constraint
length 15 convolutional decoder," 
\emph{IEEE Trans.\ Commun.}, vol.\  40, pp.\ 1810--1819, Dec.\ 1992.


\bibitem{CHIW98}
D. J. Costello, Jr., J. Hagenauer, H. Imai and S. B. Wicker,
``Applications of error-control coding,"  
\emph{IEEE Trans.\ Inform.\ Theory}, vol.\ 44, pp.\ 2531--2560, Oct.\
1998.

\bibitem{F67}
G. D. Forney, Jr., ``Review of random tree codes," Appendix A, Final
Report, Contract NAS2-3637, NASA CR73176, NASA Ames Res.\ Ctr.,
Moffett Field, CA, Dec.\ 1967.

\bibitem{F70}
G. D. Forney, Jr., ``Coding and its application in space communications," 
\emph{IEEE Spectrum}, vol.\ 7, pp. 47--58, 1970.

\bibitem{F72}
G. D. Forney, Jr., ``Maximum-likelihood sequence estimation of digital
sequences in the presence of intersymbol interference,"  \emph{IEEE Trans.\
Inform.\ Theory}, vol.\ IT--18, pp.\ 363--378, May 1972.

\bibitem{F73}  
G. D. Forney, Jr.,  ``The Viterbi algorithm,"  \emph{Proc.\ IEEE}, vol.\ 
61, pp.\ 268--278, March 1973.

\bibitem{F74}
G. D. Forney, Jr., ``Convolutional codes II.  Maximum-likelihood
decoding,"  \emph{Inform.\ and Control}, vol.\ 25, pp.\ 222--266, 1974.


\bibitem{FBEM96}
G. D. Forney, Jr., L. Brown, M. V. Eyuboglu, and J. L. Moran III, 
``The V.34 high-speed modem standard,"  \emph{IEEE Commun. Mag.}, vol.\ 34,
no.\ 12, pp.\ 28-33, Dec.\ 1996.


\bibitem{G63}
R. G. Gallager, \emph{Low-Density Parity-Check Codes}.  Cambridge, MA:  MIT
Press, 1963.

\bibitem{H68}
J. A. Heller, ``Short constraint length convolutional codes,"  Jet Prop.\
Lab., Space Prog.\ Summary 37--54, vol.\ III, pp.\ 171--177, 1968.

\bibitem{H69}
J. A. Heller, ``Improved performance of short constraint length
convolutional codes,"  Jet Prop.\ Lab., Space Prog.\ Summary 37--56, vol.\
III, pp.\ 83--84, 1969.

\bibitem{HJ71}
J. A. Heller and I. M. Jacobs, ``Viterbi decoding for satellite and
space communication,"  \emph{IEEE Trans.\ Commun.\ Tech.}, vol.\  COM--19,
pp.\ 835--848, Oct.\ 1971.

\bibitem{HSF97}
J. Henderson, S. Salzberg and K. H. Fasman,
``Finding genes in DNA with a hidden Markov model,"
\emph{J. Comput. Biol.}, vol.\ 4, pp.\ 127--141, 1997.

\bibitem{ISW98}
K. A. S. Immink, P. H. Siegel and J. K. Wolf,
``Codes for digital recorders,"  
\emph{IEEE Trans.\ Inform.\ Theory}, vol.\  44, pp.\ 2260--2299, Oct.\
1998.

\bibitem{K71a}
H. Kobayashi, ``Application of probabilistic decoding to digital magnetic
recording systems,"  \emph{IBM J. Res.\ Dev.}, vol.\ 15, pp.\ pp.\ 64--74,
Jan.\ 1971.

\bibitem{K71b}
H. Kobayashi, ``Correlative level coding and maximum likelihood decoding,"  
\emph{IEEE Trans.\ Inform.\ Theory}, vol.\  IT--17, pp.\ 586--594, Sept.\
1971.

\bibitem{KFL01}
F. R. Kschischang, B. J. Frey and H.-A. Loeliger, 
``Factor graphs and the sum-product algorithm,"
\emph{IEEE Trans.\ Inform.\ Theory}, vol.\ 47, pp.\ 498--519, Feb.\ 2001.

\bibitem{M63}
J. L. Massey, \emph{Threshold Decoding}.  Cambridge, MA:  MIT Press, 1963.

\bibitem{M99}
D. Morton, ``Andrew Viterbi, electrical engineer:  An oral history,"  IEEE
History Center, Rutgers U., New Brunswick, NJ, Oct.\ 1999.

\bibitem{O69}
J. K. Omura,
``On the Viterbi decoding algorithm,"  \emph{IEEE Trans.\
Inform.\ Theory}, vol.\ IT--15, pp.\ 177--179, 1969.

\bibitem{O71}
J. K. Omura,
``Optimal receiver design for convolutional codes and channels with
memory via control theoretical concepts," 
\emph{Info.\ Sci.}, vol.\ 3, pp.\ 243--266, July 1971.

\bibitem{P05}
R. Padovani, ``Ten years of progress in CDMA," Viterbi Conference, Univ.\
So.\ Calif., Los Angeles, Mar.\ 2005.

\bibitem{R89}
L. R. Rabiner,
``A tutorial on hidden Markov models and selected applications in speech
recognition," \emph{Proc.\ IEEE}, vol.\ 77, pp.\ 257-286, Feb.\ 1989.

\bibitem{T81}
R. M. Tanner, ``A recursive approach to low complexity codes,"
\emph{IEEE Trans.\ Inform.\ Theory}, vol.\ IT--27, pp.\ 533--547,
Sept.\ 1981.

\bibitem{U71}
G. Ungerboeck, ``Nonlinear equalization of binary signals in Gaussian
noise,"
  \emph{IEEE Trans.\ Commun.\ Tech.}, vol.\  COM--19, pp.\ 1128--1137,
Dec.\ 1971.

\bibitem{U82}
G. Ungerboeck, ``Channel coding with multilevel/phase signals,"
  \emph{IEEE Trans.\ Inform.\ Theory}, vol.\ IT--28, pp.\ 55--67, Jan.\
1982.


\bibitem{V67a}  
A. J. Viterbi,  ``Error bounds for convolutional codes and an
asymptotically optimum decoding algorithm,"  \emph{IEEE
Trans.\ Inform.\ Theory}, vol.\  IT--13, pp.\ 260--269, April 1967.


\bibitem{V70}  
A. J. Viterbi,  ``Review of \emph{Statistical Theory of Signal
Detection} (2nd ed.), by Carl W. Helstrom," 
\emph{IEEE Trans.\ Inform.\ Theory}, vol.\ IT--16, p.\ 653, Sept.\
1970.

\bibitem{V71}  
A. J. Viterbi,  ``Convolutional codes and their performance in
communication systems,"  
\emph{IEEE Trans.\ Commun.\ Tech.}, vol.\  COM--19, pp.\ 751--772,
Oct.\ 1971.

\bibitem{V90}  
A. J. Viterbi,  ``From proof to product," 1990 IEEE Communication Theory 
Workshop, Ojai, CA, April 1990.


\bibitem{W96}
N. Wiberg, ``Codes and decoding on general graphs,"
Ph.D.\ dissertation, Link\"{o}ping U., Link\"{o}ping, Sweden, 1996.

\bibitem{WLK95}
N. Wiberg, H.-A.\ Loeliger and R. K\"{o}tter, ``Codes and iterative
decoding on general graphs," \emph{Eur.\ Trans.\ Telecomm.}, vol.\ 6, pp.\
513--525, Sept./Oct.\ 1995.

\bibitem{WR61}
J. M. Wozencraft and B. Reiffen, \emph{Sequential Decoding}.
  Cambridge, MA:  MIT Press, 1961.

\bibitem{Y64}
H. Yudkin, ``Channel state testing in information decoding,"  Sc.D.\
dissertation, Dept.\ Elec.\ Engg., MIT, Cambridge, MA, 1964.

\end{thebibliography}
}
\end{document}